\documentclass[aps,preprintnumbers,eqsecnum,amsmath,amssymb,showpacs,nofootinbib,twocolumn]{revtex4}
\usepackage{bm}% bold math

\usepackage{graphics,bbm,mathrsfs}
\usepackage{color,cancel,array}

\definecolor{ossaorange}{rgb}{1, 0.9, 0.7}

\definecolor{lgray}{gray}{0.9} 		%background of the algorithm boxes
\def\softness{0.4}
\definecolor{softred}{rgb}{1,\softness,\softness}
\definecolor{softgreen}{rgb}{\softness,1,\softness}
\definecolor{softblue}{rgb}{\softness,\softness,1}
\definecolor{deepred}{rgb}{0.54, 0.18, 0.067}
\definecolor{deepgreen}{rgb}{0.141, 0.53, 0.133}
\definecolor{deepblue}{rgb}{0.3, 0.35, 1}
\newcommand*{\tl}[0]{\succ}                          % total ordering of term
\newcommand*{\hs}{\hspace}
\begin{document}
\newcommand{\locsection}[1]{\setcounter{equation}{0}\section{#1}}
\renewcommand{\theequation}{\thesection.\arabic{equation}}
\def\F{{\bf F}}
\def\A{{\bf A}}
\def\J{{\bf J}}
\def\af{{\bf \alpha}}
\def\beqn{\begin{eqnarray}}
\def\eeqn{\end{eqnarray}}

\def\dspace{\baselineskip = .30in}
\def\beq{\begin{equation}}
\def\eeq{\end{equation}}
\def\bw{\begin{widetext}}
\def\ew{\end{widetext}}
\def\pl{\partial}
\def\na{\nabla}
\def\al{\alpha}
\def\bt{\beta}
\def\Ga{\Gamma}
\def\ga{\gamma}
\def\de{\delta}
\def\De{\Delta}
\def\da{\dagger}
\def\ka{\kappa}
\def\si{\sigma}
\def\Si{\Sigma}
\def\te{\theta}
\def\La{\Lambda}
\def\lam{\lambda}
\def\Om{\Omega}
\def\om{\omega}
\def\ep{\epsilon}
\def\non{\nonumber}
\def\sq{\sqrt}
\def\sqg{\sqrt{G}}
\def\sp{\supset}
\def\sb{\subset}
\def\l{\left (}
\def\r{\right )}
\def\lq{\left [}
\def\rq{\right ]}
\def\fr{\frac}
\def\la{\label}
\def\hs{\hspace}
\def\vs{\vspace}
\def\inf{\infty}
\def\ran{\rangle}
\def\lan{\langle}
\def\ov{\overline}
\def\tl{\tilde}
\def\tm{\times}
\def\lrar{\leftrightarrow}

%\begin{document}

%\preprint{HD-THEP-08-08}

%\preprint{OSU-HEP-08-01}

%\preprint{February 6, 2008}

\vs{1cm}

\title{
The mass of the Higgs boson in the trinification subgroup of E6}
% Force line breaks with \\

\author{Berthold Stech}
\email{B.Stech@ThPhys.Uni-Heidelberg.DE}

\affiliation{Institut f\"ur Theoretische Physik, Philosophenweg 16, D-69120 Heidelberg, Germany}

\vs{1cm}

%\date{\today}% It is always \today, today,
             %  but any date may be explicitly specified

\begin{abstract}
The extension of the standard model to  $SU(3)_L\times SU(3)_R \times SU(3)_C$  is considered. Spontaneous symmetry breaking requires two Higgs field multiplets with a strong hierarchical structure of vacuum expectation values. These vacuum expectation values, some of them known from experiment, are used to construct invariant potentials in form of a sum of individual potentials relevant at the weak scale.  As in a previous suggestion \cite{BSHiggs} one may normalize the most important individual potentials such that their mass eigenvalues agree with their very large vacuum expectation values. In this case (for a wide class of parameters) the  scalar field corresponding to the standard model Higgs turns out to have the precise mass value $m_{Higgs}= \frac{ v}{\sqrt{2}} = 123 $~GeV  at the weak scale . The physical mass (pole mass) is larger and found to be $125 \pm 1.4$~GeV.
\end{abstract}
                             % Classification Scheme.
%\keywords{Suggested keywords}%Use showkeys class option if keyword
                              %display desired
\maketitle

\section{Introduction}\label{sec:1}
\vspace{-0.2cm}
\hspace{6cm}
\it {ol' man river}
\vspace{.3cm}
\rm

The imbedding of the standard model into a larger group implies an extended Higgs structure. The corresponding potential should give rise to spontaneous symmetry breaking leading to a large spectrum of  scalar boson masses. The aim of this article is to consider the structure one encounters by extending  the standard model to $E6$ \cite{E6} or rather to its maximal subgroup  $SU(3)_L\times SU(3)_R \times SU(3)_C $ . This group has to be  combined with the discrete group $Z_3$ exchanging left, right and color symmetries \cite{trini,Glashow,Babu,Willenbrock,Sayre,Pas}. It has been named trinification group by the author of Ref. \cite{Glashow}. The subgroup with  $Z_2$ (exchanging left and right) can appear as an intermediate symmetry at and above the scale where the gauge couplings $g_1$ and $g_2$ unite \cite{BZ}. 
In this model one has to deal with masses and vacuum expectation values of scalar fields which show an extreme hierarchical structure extending over many orders of magnitudes. To break the  group down to the standard model two scalar matrix fields are necessary $H$ and $\tilde H$. These two $ 3 \times 3 $ matrices of fields transform  according to $ (3^*, 3)$ with respect to $SU(3)_L$  and $SU(3)_R $  \cite{BZ} \cite{BS}. Thus, we have to deal with 36 real scalar fields. With respect to the $SU(2)_L $ subgroup they form six complex Higgs doublets and six complex singlets.  

The vacuum expectation value (vev) of the first matrix ($H$), which couples directly to the fermions,  can be chosen diagonal. Its structure is known from experiment: The element $ \lan H\ran^1_1$ describes the vacuum expectation value $ v=174$ ~GeV of the conventional Higgs field. The $ \lan H\ran^2_2 $ element is very small and of the order of the mass of the $b$ quark. For simplicity we will set it to zero here. The   $ \lan H\ran^3_3 $  element, on the other hand, is huge and its value $M$ is expected to be close to the scale of electroweak unification, i.e. the meeting point of the gauge couplings $g_1$ and $g_2$  which occurs at $\approx 2 \cdot 10^{13}$~ GeV. 

The second matrix of scalar fields $\tilde H$ is not directly coupled to fermions \cite{BZ} \cite{BS}. It needs to have a sizable vacuum expectation value for the element 
$ \lan \tilde H\ran^3_2 $ which we denote by $\tilde M$ \cite{Babu,BZ}. This vev determines the masses of the right-handed vector bosons . The experimental limit on right-handed currents provides for a lower limit for $\tilde M$  of a few $TeV$.
Already the nonvanishing of $ M $ and $\tilde M$ is sufficient to break the trinification group down to the Glashow-Weinberg-Salam group. Finally, the finite value for $v$  leaves only the electromagnetic $U(1)_{e}$ symmetry.

In a previous article, the fields of the matrix $H$ have been discussed leaving the matrix $\tilde H$ aside  \cite{BSHiggs}. The required vacuum expectation values of these fields could be obtained with the help of logarithmic contributions for the $SU(3)_L \times SU(3)_R $ invariant potentials. The input vacuum expectation values mentioned above provide for scalar masses and the 15 Goldstone bosons.  The article  contains in addition the suggestion for a speculative normalization of the potentials. This hypothesis led to a prediction for the mass of the Higgs meson of the standard model,  namely to 
 $m_{Higgs} = v/\sqrt {2} = 123$ ~GeV \cite{BSHiggs}. After the more recent experimental indications for a Higgs-like structure in the $ 125$~ GeV region \cite{Exp}  this prediction may have some significance. Therefore a more detailed discussion is indicated. It should deal with all 36 fields and should show  - in an example - the full spectrum of now $36 -15 = 21 $ scalars.  
 
Invariants which combine $H$ and $\tilde H$ fields together now play an essential role. Without those invariants several charged and neutral scalars would remain massless. In general the inclusion of such invariants will change the previously obtained mass values from the single Higgs field $H$. However, as we will see, it turns  out that for the special field, which corresponds to the standard model Higgs, no noticeable changes occur for a large class of parameters. Thus, the Higgs mass close to $123$~ GeV  can still be the consequence of the following assumption \cite{BSHiggs}: the fields in $H$ are determined by two separate potentials according to the two relevant invariants. Each potential is normalized such that the mass obtained from it coincides with the input vacuum expectation value to leading order in $M$. As a motivation we note that only with this assumption the difference between vacuum expectation value and mass eigenvalue remains finite in the large $M$ limit.

At present it seems too difficult, at least for the author, to obtain a potential valid for the high and the low scale regions which provides for spontaneous symmetry breaking of the trinification group, fulfilling thereby all theoretical requirements for this multi-Higgs problem. However, for the region of the weak scale,  effective potentials can be constructed  which give some insight into the structure of this Higgs system.  Coming down from high scales which involve very large masses, the effective potential at low scales is certainly expected to include logarithmic terms of the same order as the $H^4$ and $\tilde H^4$ terms. In fact, only the inclusion of logarithmic terms allow the potentials to have their minimum at the wanted places. Shifting the appropriate fields according to their vevs and neglecting terms with inverse powers of $M$, one gets potentials of the tree type.  No powers of fields higher than 4 appear anymore and the potential minimum remains unchanged. Moreover, the important terms containing $M$ and $ \tilde M$ occur with shifted and unshifted fields of lower powers. This way the task is simplified and one gets, at low scales, a model only slightly different from the standard model.  As we will see, the potential constructed leads to the required spontaneous symmetry breaking. All Higgs fields obtain masses in agreement with experimental bounds.

\section{ Scalar Fields with low and very high scale vacuum expectation values in
the same representation.}\label{sec:2}

An interesting example of a scalar Higgs field in a grand unified model is the field $ H_{27} $, the irreducible "27" representation of $ E6$. In a nonsupersymmetric version of $E6$ \cite{BZ} \cite{BS} the maximal subgroup of $E6$, the symmetry $ SU(3)_L \times SU(3)_R \times SU(3)_C $  (the trinification group) plays an important role together with the  symmetry $Z_3$ which allows the exchange of "left" with "right" and "color". Apart from the breaking $Z_3 \rightarrow Z_2$, this symmetry holds from the point of electroweak unification up to the complete gauge group unification .  In this region which starts at $\approx 2 \cdot 10^{13}$~ GeV the gauge couplings $g^n_1= \sqrt{5/3}~ g_1$ and $g_2 $ are identical. 

Let us consider for the moment the color singlet part of a single $H_{27}$ multiplet field only, neglecting the influence of the other fields. It is a $ 3 \times 3 $ matrix field $H^i_k $  where the index $i$ transforms as an upper index with respect to $ SU(3)_L $ while the lower index $k$ transforms according to $ SU(3)_R$. The indices $i=1,2$ are the $SU(2)_L$ indices of the standard model. $H^i_k $  contains 18 real fields:

The vacuum expectation value of $H$ can be chosen to be a diagonal matrix with real and positive elements by absorbing transformation matrices  and phases by the fermion fields. In the following we use $H = h+i f$  where $h$ and $f$ are matrices of real fields. 
\begin{eqnarray}
\lan H \ran ~ = ~\lan h \ran ~~~ ~~= 
\hspace{0.0cm}
\begin{array}{ccc}
& {\begin{array}{ccc}
 & &
\end{array}}\\ \vspace{2mm}
~~~~~~~~ \hs{-0.5cm}
\begin{array}{c}
  \\
\end{array} \hspace{-0.1cm}&{\left(\begin{array}{ccc}
v & 0& 0
\\
0 & b & 0
\\
0 & 0 & M
\end{array}\right)~.
}
\end{array}
\end{eqnarray}

Phenomenology tells us the values of $v$ and $b$: $v$ can be identified with the vacuum expectation value of the Higgs field of the standard model, $v = 174$~GeV. This element couples to the top quark. $b$ couples to the bottom quark and is much smaller, roughly equal to the mass of the bottom quark at the  weak scale. We will set $b=0$ for simplicity. The value $M$ on the other hand must be huge. It is the mass of the heavy down quark $D$, which is likely to be of the order of the scale where the $SU(3)_L \times SU(3)_R \times Z_2 $ symmetry sets in  \cite{BZ}: $\langle H \rangle^3_3 = M \simeq m_D \approx 2 \cdot 10^{13}$~GeV .

For the fields in $H$, only two $SU(3)_L \times SU(3)_R$ invariants with nonvanishing vevs exist:
\begin{eqnarray}
\label{J}
J_1 = (Tr[H^\dagger \cdot H])^2, \quad %\text{and}  \quad 
J_2  = 
Tr[ H^\dagger  \cdot H \cdot H^\dagger \cdot H].\nonumber\\
\end{eqnarray}
Their vevs are $\langle J_1 \rangle=(M^2+v^2)^2 $ and $\langle J_2 \rangle = M^4+v^4$, respectively.

It is easily seen that a tree potential with the two invariants (\ref{J}) cannot produce the wanted vacuum expectation values. To have  a potential with the required properties at the weak scale the logarithmic dependence on $J_1$ and $J_2$ must be included. In contrast to the usually very small Coleman-Weinberg term \cite{CW} one expects here the log term to be of the same order as the tree part because of the extremely long way down from the very heavy states. Indeed, large log terms follow necessarily:  Our requirements are satisfied by a linear combination of the two separate potentials:
\begin{eqnarray}
\label{VJ}
V_1 &=&  \frac{1}{8}~J_1~\left(\log{\frac{J_1}{\langle J_1 \rangle}} - 1\right)\quad  \text{and} 
\nonumber
\\
V_2&=&  \frac{1}{8}  ~J_2~ \left(\log{\frac{J_2}{\langle J_2 \rangle}} - 1\right) .
\end{eqnarray}
Higher powers of the log terms can also be used.  For instance
\begin{eqnarray}
\label{VJ1}
V_1 =\frac{1}{8}~\frac{J_1}{1+\log{\frac{J_1}{\langle J_1 \rangle}}} \quad  \text{and} \quad 
V_2= \frac{1}{8} ~\frac{J_2}{1+\log{\frac{J_2}{\langle J_2 \rangle}}}.\nonumber\\
\end{eqnarray}
Equations (\ref{VJ}) and (\ref{VJ1}) are equivalent in our treatment. 
First and second derivatives at the minimum are the same and thus lead to the same scalar particle spectrum.
The scalar potential taken as a linear combination of $V_1$ and $V_2$ (with positive coefficients) has the wanted properties. This potential is fully invariant and  provides for the spontaneous symmetry breaking to $U(1)\times U(1)_e$. The derivatives of this potential with respect to the 18 fields vanish at the minimum and the $18 \times 18$ matrix for the second derivatives at the minimum leads to two massive states, one massless state and 15 Goldstone bosons. In Eqs(\ref{VJ}, \ref{VJ1}) the  factors  $1/8$
 serve as a
normalization of the potentials. They are chosen such that --to order $M$ -- the mass values obtained from the second derivatives coincide with the input vacuum expectation values according to our postulate.  This postulate insures that the difference between mass and vev is not proportional to the huge value $M$. It remains finite in the limit $M \rightarrow \infty $. The normalization will be of significance as shown below. 

The potentials obtained so far cannot be used to calculate radiative corrections.  One has to extract   a  tree potential which can then be improved by renormalization group methods in the conventional way. To do this an expansion in $M$ is necessary. It can be done after  the shift of the fields according to their vevs ($h^1_1  \rightarrow  v + h^1_1$, $h^3_3 \rightarrow  M + h^3_3 $ ) is performed. By neglecting inverse powers of $M$, the above potentials then become polynomials with field configurations up to the fourth power only.  Moreover, the interesting terms  proportional to $M^2$,  $v\cdot M$  and $v^2$ occur  with fields to second order only. The symmetry breaking properties remain unchanged. The relevant scale for this potential is the weak scale.
 
\begin{eqnarray}
\label{VM}
V_1 &=&  (h^3_3)^2 M^2 + 2 h^1_1 h^3_3~ v M + (h^1_1)^2~ v^2 \nonumber\\
&&+ O(H^3) + O(H^4)~, \nonumber
\\     
V_2  &=&  (h^3_3)^2 M^2  + O(H^3) + O(H^4) ~.
\end{eqnarray}

Constant terms are left out and the field combinations of third and fourth order are not shown explicitly. They are not relevant at the minimum but are needed for renormalization. $V_1$ and $V_2$ of Eq.(\ref{VM}) can now be used as the tree potential at the weak scale. It replaces the standard model potential in the presence of a huge hierarchy. The  Higgs field $h^1_1$  and the $SU(2)_L$ singlet field $h^3_3$ are both real fields.
As in \cite{BSHiggs} we make the assumption to use the two potentials with equal weight because of their identity in the limit $v \rightarrow 0$. The normalization is decisive and kept.
\beq
\label{specV}
V_{tree} =  V_1 + V_2 .
\eeq
This potential has a minimum for zero values of all (shifted) fields. The two massive states have the masses
(shown to order $v^2$)
\beq
\label{H123}  m_{Higgs}^2  = \frac{1}{2}  v^2 \qquad \text{and} \qquad M_H^2 = 2 M^2 +\frac{1}{2} v^2
\eeq            
Thus, the prediction for the Higgs mass at the weak scale is as 
in \cite{BSHiggs}
\beq
\label{mH}
 m_{Higgs} = \frac{1}{\sqrt{2}}~v \simeq 123~ GeV .
\eeq
However, the influence of the fields of the second Higgs multiplet $\tilde H$ can change this result. 
This is the subject of the next section together with the attempt to obtain the full mass spectrum of all scalars.

\section{ Two $SU(3)_L \times SU(3)_R$  scalar field multiplets.\label{sec:3 }}
The second multiplet of scalar fields $\tilde H$ is required from phenomenological considerations. In the model \cite{BZ} it is not directly coupled to the fermions, only via gauge vector bosons. It can have its own paritylike symmetry: $\tilde H \rightarrow - \tilde H $. Its vevs cannot  be diagonalized anymore when keeping $\langle H \rangle $ diagonal.  For the breaking of the original left-right symmetry,
the dominant vev has to be at the $(3,2)$ position \cite{Babu}.
Defining $\tilde H = \tilde h + i \tilde f$, one has
 \begin{eqnarray}
 \lan \tilde H \ran ~ = ~\lan \tilde h \ran ~~~ ~~= 
 \hspace{0.0cm}
\begin{array}{ccc}
& {\begin{array}{ccc}
 & &
\end{array}}\\ \vspace{2mm}
~~~~~~~~ \hs{-0.5cm}
\begin{array}{c}
  \\
\end{array} \hspace{-0.1cm}&{\left(\begin{array}{ccc}
0 & 0& 0
\\
0 & 0 & 0
\\
0 & \tilde M & 0
\end{array}\right)~.
}
\end{array}
\end{eqnarray}
The value of $\tilde M$ fixes the masses $m_{W_R}$ of new vector bosons coupled to right-hand vector currents:  $m_{W_R}/ m_{W_L}  \simeq  \tilde M / v$. These right- hand currents are, apart from their helicity structure, of the same form as the well- known left-hand currents. LHC experiments can discover these bosons if their masses lie in the TeV region. In the following we take $ v \ll \tilde M \ll  M$.

The wanted total potential should now provide masses for all fields. One needs new invariants depending on $\tilde H$ but necessarily also invariants combining $ H$ and $\tilde H$ fields. Otherwise one would have additional symmetries and thus many massless scalar particles.                                                                 
Important invariants which also respect the symmetry $\tilde H \rightarrow - \tilde H $ are
\begin{eqnarray}
\label{JJ}
J_3 &=& Tr[\tilde H^\dagger \cdot \tilde H]^2 , \nonumber\\
J_4  &=& 
Tr[ \tilde H^\dagger  \cdot \tilde H \cdot \tilde H^\dagger \cdot \tilde H] ,  \nonumber\\
J_5&=& Tr[\tilde H^\dagger \cdot \tilde H \cdot H^\dagger \cdot H] , \nonumber\\
J_6 & =& Tr[ \tilde H^\dagger  \cdot H \cdot H^\dagger \cdot \tilde H].
\end{eqnarray}
Their vevs are $\lan J_3 \ran = \lan J_4 \ran = \tilde M^4 , ~\lan J_5 \ran = 0, 
~\lan J_6 \ran = \tilde M^2  M^2$.

The invariant $J_5$ can be directly added to the potential (\ref{specV}) since all its $36$ first derivatives vanish at the proposed minimum.  $ J_6 $, on the other hand, has to be combined with $J_3$ or $J_4$:
the first derivatives of  $J_4 - 2 \tilde M^2/ M^2 J_6$ are not strictly zero at the minimum but vanish for large $M$. Thus this combination can be used. Therefore, a suitable and still simple tree potential for the $36$ scalar fields reads: 
\begin{eqnarray}
\label{fpot}
V_{tree} &=&  2 (h^3_3)^2 M^2 + 2 h^1_1 h^3_3 v M + (h^1_1)^2 v^2  \nonumber
\\
&&
+  r_4 (J_4 - 2 \frac{ \tilde M^2}{ M^2}  J_6) + r_5  J_5 
\nonumber\\
&&
+ O((H, \tilde H)^3) + O((H, \tilde H)^4).
\end{eqnarray}
In this equation the shifted fields (also $\tilde h^3_2 \rightarrow \tilde M + \tilde h^3_2$) have to be used and, after expanding in powers of $M$, the inverse powers of $M$ have to be neglected. Similar to Eq. (\ref{specV}), the combination of fields not contributing to the tree level spectrum of particles are not shown explicitly. They are, however, important for the scale dependence. The coefficients $r_4$ 
and $r_5$ have to be positive and not  too small but are otherwise arbitrary. The formula (\ref{fpot}) is simple enough to obtain  analytically the eigenvalues and eigenvectors of the $36 \times 36$ matrix of second derivatives. The square of the masses are 
\begin{eqnarray}
\label{masses}
&&\frac{v^2}{2},~~2 M^2 +\frac{v^2}{2}, ~~(M^2 +\tilde M^2) r_5~  (2\times), ~~ M^2 r_5 ~(4\times),
\nonumber\\
&&4 \tilde M^2 r_4 ,~~ (\tilde M^2+v^2) r_5~(2\times) ,
~~
2\tilde M^2 r_4~(4\times) ,
\nonumber\\
&&\tilde M^2 r_5 ~(2\times), ~~v^2 r_5 ~(4\times),~~
 ~~0 ~(15 \times).
\end{eqnarray}
It is seen that the first and the second term are unchanged as compared to Sec. 2. The $15$ zero mass Goldstone particles will become the longitudinal parts of the gauge vector bosons. The other masses depend on the parameters $r_4$, $r_5$, $M$ and $\tilde M$. Interestingly, there is one Higgs-like multiplet independent of $M$ and $\tilde M$ depending only on $r_5$ and $v$. It will be referred to as the ``second Higgs''. The diagonalization of the $36 \times 36$ matrix allows to identify all fields. After taking account of tiny mixings, the new field $h^1_1$  describes the field of the standard model-like Higgs and the second Higgs consists of the (also slightly changed) four fields $\tilde h^1_1, \tilde f^1_1, \tilde h^2_1, \tilde f^2_1$ from the matrix $\tilde M$. 
One can add to the potential the invariant  $J_7 = Tr[\tilde H^\dagger  H]~ Tr [ H^\dagger \tilde H]$  and    also  $J_3$ in $\log$ form similar to $V_1$.  These contributions do not affect the value of the Higgs mass. Thus, for a wide class of potentials the result for the Higgs obtained here and in \cite{BSHiggs} can remain valid.

But certainly,  it is also possible to devise special potentials which have quite different properties. There is then less connection with the standard model.

\section{The low scale region.\label{sec:4}}

Let us now look at the low scale region near the mass of the Higgs. 
One can start from Eq.(\ref{fpot}) which, if written in detail, contains a large amount of 
different field configurations. However, drastic simplifications occur after diagonalizing 
the $36 \times 36 $ mass matrix and going with $M$ and $\tilde M$ to infinity. 
This allows us to neglect $16$ heavy fields (the $16$ masses are given in Eq.(\ref{masses})). 
The rest of the fields consist of the ``second Higgs'' (four states), which may or may not 
be very heavy compared to the Higgs,  and finally the Higgs field  $h^1_1$ itself and the $15$ 
massless Goldstone bosons. The latter become the longitudinal components of the gauge bosons. 
In the low scale limit only the coupling to the top and to the three low mass vector 
bosons $W^+,W^-,~ Z$ need to be considered.
 
The relevant part of the Lagrangian for this low scale domain in the unitary gauge turns out to be:
\begin{eqnarray}
\label{field coupling} 
{\cal{L}}_{h} &=& - \frac{1}{4} h^4 - ~\frac{1}{2} ~v ~h^3 - ~\frac{1}{2} ~v^2 ~h^2+ g_t ~\bar t t~ h 
\nonumber \\
&&+ (g_2)^2~ W^+ W^- v~ h +\frac{1}{2} \l(g_1)^2 + (g_2)^2\r~ Z^2 ~v ~h
\nonumber\\
&& - r_4  \l |\tilde H_1^1|^2 +|\tilde H_2^1|^2 \r^2 
\nonumber\\
&&
-  r_5  (h^2 + 2 h v+ v^2  ) 
\l |\tilde H_1^1|^2 +|\tilde H_2^1|^2\r .
\end{eqnarray}
For convenience the field $h^1_1$ is denoted by $h$. According to our derivation the sum of the first three terms in (\ref{field coupling}) can be identified with the renormalized potential  for the Higgs field at the weak scale. The next three terms describe the coupling of $h$ to the  top and to the vector bosons $W^+,~W^-,~ Z$. The final terms 
contain the fields of the second Higgs which need only be considered if their masses are relatively low.
Since these fields have no vev and are not directly coupled to fermions, their neutral members are possible candidates for dark matter particles.

It is seen that our model has the form of the standard model except for the second Higgs and the larger number of Goldstone bosons. The ratio between the $h^3$ coupling constant and the mass term  is the same as in the standard model. Only the $h^4$ coupling is a factor two larger.

Introducing the scale dependent coupling coefficient $\lambda (\mu)$  Eq.(\ref{field coupling}) leads to the boundary  condition  $ \lambda(\mu_0)= 1/8 $. Here, $\mu_0$ stands for the weak scale at which the matching with $h$ should be performed. It seems appropriate to take for this scale the sum of the square of the four boson masses caused by the Higgs particle: $\mu_0^2 = m_H^2 + 2 m_W^2 + m_Z^2  \simeq (192~GeV)^2$. The  Higgs mass and $\lambda$ are related according to $ m_{Higgs}^2 = 4  \lambda(\mu)  v^2 $ . 

To get the physical mass of the Higgs (the pole mass) in our model is clearly of interest. This requires us to study the scale-dependence of the mass in the low scale region. A systematic analysis of Eq.(\ref{field coupling}) is required here but has not been done. However, one can get an estimate by using as an approximation the standard model result for the
connection between the scale-dependent $\overline{MS}$  coupling constant $\lambda(\mu)$ and the pole mass according to Ref.\cite{Sirlin}: 
\beq
\label{Sirlin}
m_{Higgs}^2|_{pole} =  \frac{\lambda(\mu)}{1 + \delta (\mu)}~4 ~v^2 .
\eeq
The scale-dependent correction term $\delta (\mu) $ can be calculated from particle couplings and masses \cite{Sirlin}. We use in this expression for the mass of the top quark $m_t = 173~GeV$ and for the Higgs mass $123 ~GeV$. 

 An estimate of the pole mass can now be obtained by fixing $\lambda(\mu_0 = 192~ GeV)=1/8$. Applying the renormalization group equations of the standard model and taking $\mu$ in the interval $110~GeV < \mu <   250~GeV$ one finds from (\ref{Sirlin}) 
$$ m_{Higgs}|_{pole} ~ = ~125  \pm 1.4 ~GeV. $$ 

%{\bf Acknowledgment}\\
\acknowledgments
It is a pleasure to thank my colleagues Jan Pawlowski, Tilman Plehn,
 Michael Schmidt and Werner Wetzel
for stimulating discussions.
%\vs{0.2cm}

\end{document}